\documentclass[%
 reprint,
 amsmath,amssymb,
 aps,
pre]{revtex4-2}

 \usepackage{amssymb}
 \usepackage{amsmath}
 \usepackage{mathtools}
 \usepackage[T1]{fontenc}
 \usepackage[utf8]{inputenc}
 \usepackage{color,xcolor}
 \usepackage{esvect}

\definecolor{cadmiumgreen}{rgb}{0.0, 0.42,0.24}

\newcommand{\pa}{\partial}

\newcommand{\myref}[1]{(\ref{#1})}

\newcommand{\om}{\omega} 

\newcommand{\de}{\delta}
\newcommand{\De}{\Delta}
\newcommand{\al}{\alpha}

\newcommand{\la}{\lambda}
\newcommand{\La}{\Lambda}
\newcommand{\ga}{\gamma}

\renewcommand{\hat}[1]{\widehat{#1}}

\newcommand{\demi}{\frac{1}{2}}

\newcommand{\mcal}[1]{\mathcal{#1}}

\newlength{\somme}
\newlength{\sommep}
\newlength{\sommebis}
\newlength{\sommepbis}
\makeatletter
\newcommand{\pushright}[1]{\ifmeasuring@#1\else\omit\hfill$\displaystyle#1$\fi\ignorespaces}
\newcommand{\pushleft}[1]{\ifmeasuring@#1\else\omit$\displaystyle#1$\hfill\fi\ignorespaces}
\makeatother

\makeatletter 
\newcommand\notsotiny{\@setfontsize\notsotiny{6}{7}}
\makeatother

\usepackage{contour}
\usepackage{ulem}

\contourlength{0.8pt}

\usepackage{graphicx,siunitx,float}
\usepackage{dcolumn}
\usepackage{bm}

\begin{document}

\preprint{APS/123-QED}

\title{Elastic ribbons in bubble columns:\\ when elasticity, capillarity and gravity govern equilibrium configurations}

\author{Jean Farago$^{1}$}
\author{Manon Jouanlanne$^{1}$}
\author{Antoine Egelé$^{1}$}
\author{Aur\'elie Hourlier-Fargette$^{1}$}
\affiliation{
$^1$ Université de Strasbourg, CNRS, Institut Charles Sadron UPR22, F-67000 Strasbourg, France\\}

\date{\today}

\begin{abstract}
Taking advantage of the competition between elasticity and capillarity has proven to be an efficient way to design structures by folding, bending, or assembling elastic objects in contact with liquid interfaces. Elastocapillary effects often occur at scales where gravity does not play an important role, such as in microfabrication processes. However, the influence of gravity can become significant at the desktop scale, which is relevant for numerous situations including model experiments used to provide a fundamental physics understanding, working at easily accessible scales. We focus here on the case of elastic ribbons placed in two-dimensional bubble columns: by introducing an elastic ribbon inside the central soap films of a staircase bubble structure in a square cross-section column, the deviation from Plateau's laws (capillarity-dominated case dictating the shape of usual foams) can be quantified as a function of the rigidity of the ribbon. For long ribbons, gravity cannot be neglected. We provide a detailed theoretical analysis of the ribbon profile, taking into account capillarity, elasticity and gravity. We compute the total energy of the system and perform energy minimization under constraints, using Lagrangian mechanics. The model is then validated via a comparison with experiments with three different ribbon thicknesses.

\end{abstract}

\maketitle


\section{Introduction}

Systems involving both elastic and capillary forces have been attracting a growing interest to tackle practical challenges and provide novel engineering and design tools, compelling for the microfabrication of 2D surfaces and 3D structures \cite{DeVolder2013, Kwok2020}, for the design of actuators in the emerging soft robotics field \cite{Hines2016}, but also in the context of biomechanics or biomimetic approaches \cite{Heil2008, Gernay2016, Elettro2016}, to cite only a few cases. The field of elastocapillarity includes not only the deformation of soft solids by surface tension \cite{Style2017} but also systems where the geometry of the elastic objects is of particular importance, especially in the context of slender structures for which bending dominates over stretching \cite{Holmes2019}. In appropriate conditions, capillary forces are sufficient to force such slender structures to undergo large deformations, in systems at multiple length scales, provided that capillary and elastic energies are of the same order of magnitude \cite{Roman2010, Bico2018}. A fundamental understanding of elastocapillary phenomena can be acquired with experiments at the desktop scale \cite{Bico2004, Py2007}, for systems conceptually very close to those occurring in microfabrication processes \cite{Lau2003, Legrain2014}. However, when reaching centrimetric scales, the effects of gravity on slender elastic structures can be significant. Taking capillarity aside, a typical daily life experience of slender structures for which gravity matters is the shape of curly hair \cite{Miller2014}. A few examples involving elasticity, capillarity and gravity have been studied in the literature, for instance in the context of floating thin elastic films \cite{Huang2010}, capillary in-drop spooling \cite{Elettro2018}, elastocapillary-driven snap-through \cite{Fargette2014}, or of capillary rise between elastic sheets, revisiting Jurin's law in the context of elastocapillarity  \cite{Kim2006}.\\

Here we focus on the question of the influence of gravity on the equilibrium shape of centimetric ribbons inserted inside two dimensional bubble columns. When confined inside square cross-section tubes, bubbles rearrange themselves into well ordered structures that depend on the confinement ratio (bubble size divided by the width of the tube) \cite{Hutzler2009, Reinelt2001}. By choosing an appropriate ratio, we can obtain the so-called \textit{staircase structure} which is quasi 2D (Fig.\ref{fig:sketch}a). A ribbon can then be inserted inside the central soap films, and its shape depends on the minimisation of the total energy of the system. In our previous work \cite{Jouanlanne2022, Shishkov2022}, we placed ourselves in conditions where gravity is negligible, by considering only the very bottom of the elastic ribbons. The total energy of the system was thus composed of the bending energy of the ribbon and of the interfacial energies, and we studied how the presence of the ribbon modified the classical Plateau's laws \cite{Plateau1873} for foams. Now, we consider long ribbons for which a flattening due to gravity is observed, especially in the upper parts, due to the weight of the lower parts.

\section{\label{sec:level1} Model description and notations}

\begin{figure}[h!]
    \centering
    \includegraphics{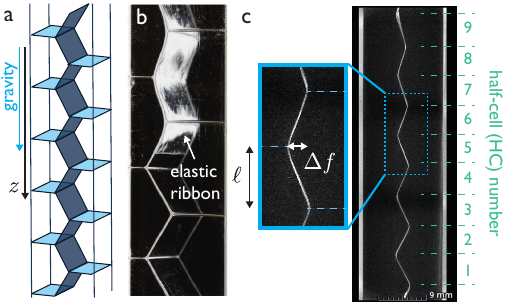}
    \caption{\textbf{Illustration of the system studied:} (a) Schematic representation of a bubble column in a tube of square cross-section. The ratio of the bubble size to the container dimension is chosen to obtain the so-called \textit{staircase structure} \cite{Hutzler2009, Reinelt2001}, invariant by translation along the axis perpendicular to the tube. (b) Experimental visualisation of the insertion of an elastic ribbon (polydimethylsiloxane (PDMS)) in a staircase bubble structure. (c) X-ray microtomography slice of the bubble column containing a thin PDMS ribbon (thickness $t=$45 $\unit{\micro\meter}$), and zoom on a half-cell presenting the definitions of the half bubble size $\ell$ and of the amplitude $\Delta f$. Note that the half-cell numbers are defined by starting from the bottom of the ribbon.}
    \label{fig:sketch}
\end{figure}

As shown in Figure  \ref{fig:sketch}, we describe a system similar to that  studied in \cite{Jouanlanne2022}. An elastic ribbon of length $L$ is inserted in the middle of a 2D staircase arrangement of bubbles of longitudinal size $2\ell$. We divide the ribbon in half-cells (HC), with numbers defined by starting from the bottom of the ribbon (HC n°1 corresponds to the first half period containing a ribbon in the whole central Plateau border). The ribbon is characterized by a bending modulus per transverse length $\al=Et^3/(12(1-\nu^2))$ where $t$ is the ribbon thickness, and ($E,\nu$) respectively the Young modulus and the Poisson ratio of the ribbon, and a lineic mass per transverse length $\la$ ($\la=\rho t$ with $\rho$ the ribbon mass density). Contrary to the case studied in \cite{Jouanlanne2022}, the gravity $g$ operating in the longitudinal $Oz$ direction is not supposed to be negligible. Its main effect is to flatten the upper regions of the ribbon due to the weight of the lower regions pulling downwards. The purpose of the present theory is to compute the equilibrium shape of a long ribbon, tied at its upper end at $z=0$, in the limit of a long ribbon $L/\ell\gg 1$ (to avoid finite size effects difficult to handle, which moreover would obscure the understanding of the theory).

The elastocapillary energy excess $\De E$ per unit (transverse) length of the system, with respect to the situation where only the bubbles are present is given by
\begin{multline}
    \De E=\frac{\al}{2}\int_{0}^L {\rm d}s\ C^2-\la g\int_0^L {\rm d}s\ z(s)\\+2\ga L-\ga\sum_{\rm HC}[\De f_i+\sqrt{3}\ell]\label{DeltaE}
\end{multline}
where $s$ is the curvilinear coordinate along the ribbon, starting from $s=0$ at $z=0$, $C=f''(z)/\sqrt{1+[f'(z)]^{3/2}}$ is the local ribbon curvature (we denote $y=f(z)$ the profile of the ribbon in the lateral direction), and $\De f_i$ is the lateral amplitude of the ribbon deformation in the $i$-th HC (we choose to number the HC from the bottom ($n=1$) upward, until $n=N$). The first three terms of the right hand side (rhs) relate to the ribbon, as respectively its bending, gravitational and  interfacial energies, the latter constant term assuming a perfect wetting of the ribbon. The second line of \myref{DeltaE} is the excess interfacial energy the system has to pay to substitute in each HC the original liquid staircase pattern by a wet ribbon : if $h$ is the width of the lateral  dimension of the tube (i.e. visible in fig. \ref{fig:sketch} c), this energy difference in the $i$-th HC is indeed $\ga[2L+h-\De f_i-(\frac{4\ell}{\sqrt{3}}+h-\frac{\ell}{\sqrt{3}})]=2\ga L-\ga[\De f_i+\sqrt{3}\ell]$. The $\sqrt{3}\ell$ term must not be discarded as a constant, due to the fact that the number of HC is variable in the optimization process.

\medskip

In writing \myref{DeltaE}, one assumed that the HC size $\ell$ is a constant along the tube, moreover unaffected by the presence of the ribbon. This approximation is justified in the Appendix at the end of the article. The direct minimization of \myref{DeltaE} is complicated, because the gravitational term breaks the vertical translational invariance. However the weight of the ribbon contained in just one HC is very modest, and it is relevant to assume that the gravitational energy may be considered as constant at the scale of the HC. This assumption assumes that the weight (per unit transverse length) of just one HC, $\sim \la g\ell$ is relatively small with respect to the typical forces (per unit transverse length), which can be either $\ga$ or $\al/\ell^2$ according to the regime considered (see later). This assumption amounts to replace $-\la g\int_{{\rm HC }n} {\rm d}s\ z(s)$ by $-\la g(N-n+\demi)\ell \int_{{\rm HC }n}{\rm d}s$, where the integral encompasses the part of the ribbon present in the $n$-th HC. In doing so, one assumes that the center of mass of the ribbon stays always in the middle of the HC, which is in particular provided by configurations where the ribbon shape is centrosymmetric within each HC. This would be exactly the case for the optimal ribbon shape in the $n$-th  HC if its weight was negligible, because in this case, the gravitational pulling forces exerted by the neighbouring HCs  on each end of the ribbon part contained in the HC would be strictly identical.

\medskip

For a long ribbon, a large upper amount is totally flattened by the weight of its lower part, so it is relevant to remove from $\De E$ the constant energy of a totally flattened ribbon, $(\De E)_{\rm flat}=2\ga L+\sum_{n=1}^{L/\ell}[-\la g(\frac{L}{\ell}-n+\demi)\ell^2-\ga\sqrt{3}\ell]$, and define
\begin{align}
    \De \mcal{E}&=\De E-(\De E)_{\rm flat}+\ga\La\left[\int{\rm d}s-L\right]
\end{align}
where $\La$ is a dimensionless Lagrangian multiplier added to encode the constraint that the total length of the ribbon is conserved.
If $N$ denotes the total number of HC the ribbons spans, the preceding expression can be recast into :
\begin{align}
    \De\mcal{E}&=\sum_{n=1}^N \left[\frac{\al}{2}\int_{{\rm HC}n}{\rm d}s\ C^2-\ga\De f_n\right.\nonumber\\
    &\phantom{a}\left.+[\la g \ell(n-\tfrac{1}{2})+\ga\La]\left(\int_{{\rm HC}n} \!\!\!\!\!\!\!\!{\rm d}s -\ell\right)\right]+\ga\ell\mcal{R}_N\label{pouetDE}\\
    \mcal{R}_N&=\frac{\la g\ell}{2\ga}(L/\ell-N)^2+(\sqrt{3}-\La)(L/\ell-N)\label{RN}
\end{align}
With this writing, one can expect that the bracketed series be convergent if $N\rightarrow\infty$, since the large values of ${n}$ will be cancelled by the terms $\propto(\int {\rm d}s-\ell)$, very small at the top of the ribbon (we will see that a remaining logarithmic divergence remains, however negligible in the limit $N\gg 1$). For sake of simplicity, the $L/\ell\rightarrow\infty$ (which entails $N\rightarrow\infty$) limit will be systematically considered in the following. It corresponds to taking the asymptotic sum of the series in \myref{pouetDE}, but keeping in mind that $L/\ell-N$ remains finite. If finite $N$ corrections would have to be considered, then the boundary effects associated to the extremities of the ribbon ending in general in the middle of a HC should be carefully estimated, a more detailed analysis which is beyond the scope of this work.

We make the problem dimensionless by defining
\begin{align}
    \widehat{\la}&=\frac{\la g \ell}{\ga}\\
    \eta&=\frac{\al}{\ga\ell^2}
\end{align}
and writing (note that $\De\mcal{E}$ is an energy per transverse unit length) :
\begin{align}
    \De\mcal{E}/[\ga\ell]&=\sum_{n=1}^N e_n+\mcal{R}_N\label{DeltaEad}\\
    e_n&=\frac{\eta}{2}\int_{{\rm HC}n}\frac{{\rm d}s}{\ell} (C\ell)^2+\eta\kappa_n^2\left(\int_{{\rm HC}n}\frac{{\rm d}s}{\ell}-1\right)-\frac{\De f_n}{\ell}\label{ousma}\\
    \kappa_n^2&=\frac{\widehat{\la}({n}-\demi)+\La}{\eta} \label{ousmakappa}
\end{align}
In the preceding formula, the curvature $C$ is related to the profile $f(z)$ of the ribbon by the already mentioned classical formula $C=f''(z)/\sqrt{1+[f'(z)]^{3/2}}$, and ${\rm d}s=\sqrt{df^2+dz^2}=dz\sqrt{1+[f'(z)]^2}$.
The optimization of \myref{DeltaEad} consists in (i) writing the Euler-Lagrange equations for $f(z)$ of the ribbon in each HC separately, which is possible since the optimal solution has extrema at the boundaries of the HC, (ii) Express the constraint ``ribbon length$=\!L$'' to remove $N$ from  \myref{DeltaEad} in favour of $\La$, and (iii) Extremalize the resulting expression for $\De\mcal{E}$ with respect to $\La$, which is equivalent to (but easier than) extremalizing with respect to $N$. Notice that the whole optimal solution is twice differentiable everywhere with slight discontinuities of the second derivatives at the boundaries of HC, which is not forbidden by the physics, as the lateral liquid interfaces act as localized (``impulsive'') normal forces at these boundaries on the ribbon.

In the general case, the optimization of \myref{DeltaEad} is probably intractable, but as in \cite{Jouanlanne2022}, we can separately solve the case $\eta\gg 1$ (stiff ribbons), where the ribbon remains everywhere almost flat, and $\eta\ll 1$ (soft ribbons), where the ribbon perturbs only slightly the Plateau's laws. In the latter case, a slight difficulty must be overcome, since the gravity significantly flattens the bubbles in the upper parts of a long enough ribbon, even if the ribbon elasticity does not.

\subsection{Stiff ribbon limit : $\eta\gg 1$}\label{etagrand}

This case is certainly less relevant experimentally, but it is theoretically instructive, as much simpler to proceed with. Here $f\ll \ell$ everywhere and it is relevant to  Taylor-expand $e_n$ up to the second order :
\begin{align}
    e_n&\simeq\frac{\eta \ell}{2}\int_0^{\ell} {\rm d}z [f''(z)]^2+\frac{\eta\kappa_n^2 \ell^{-1}}{2}\int_0^\ell {\rm d}z [f'(z)]^2-\frac{\De f_n}{\ell}\label{pouet}
\end{align}
(notice that a translation in $z$ is made so that $z$ spans the interval $[0,\ell]$ whatever the HC considered). One writes $f(z)=(\De f_n) g_n(z/\ell)$, since the particular structure of \myref{pouet}  allows a separate optimization of the shape $g_n(z/\ell)$ and the amplitude $\De f_n$ of the profile. Solving the 4-th order Euler-Lagrange equation  with the four boundary conditions $(g_n(0)=0,g_n(1)=1,g_n'(0)=g_n'(1)=0)$ gives for the optimal shape \footnote{As we consider only half cells (HC), one should impose $g_n(0)=1$ and $g_n(1)=0$ every two HC. But this point does not change any of the conclusions in the adopted assumption of light ribbons, so we impose $g(0)=0$ and $g(1)=1$ everywhere for sake of simplicity} :
\begin{multline}
    g_{n,\rm opt}(u=\tfrac{z}{\ell})\\=\frac{\tanh\left(\frac{\kappa_n}{2}\right)[\cosh(\kappa_n u)-1]+\kappa_nu-\sinh(\kappa_nu)}{\kappa_n-2\tanh\frac{\kappa_n}{2}}\label{gn}
\end{multline}
To get the optimal value of $\De f_n$, we express the energy $e_n$ optimized with respect to $g_n$ as
\begin{align}
    e_n
    &=\frac{\eta}{2}\left(\frac{\De f_n}{\ell}\right)^2\int_0^1{\rm d}u g'_{n,{\rm opt}}[-g^{(3)}_{n,\rm{opt}}+\kappa_n^2g'_{n,{\rm opt}}]
    -\frac{\De f_n}{\ell}
\end{align}
By optimizing $e_n$ with respect to $\De f_n$, we obtain
\begin{align}
\frac{\De f_{n,\rm opt}}{\ell}&=\eta^{-1}\frac{\kappa_n-2\tanh\frac{\kappa_n}{2}}{\kappa_n^3}\label{deltaf}\\
e_{n,\rm opt}&=-\demi\frac{\De f_{n,\rm opt}}{\ell}
\end{align}

\medskip

As announced in anticipation, it can be observed that the optimal value of the series in \myref{DeltaEad} keeps a slight divergence, since $e_{n,\rm opt}\sim -\frac{1}{2\widehat{\la}{n}}$ for large values of ${n}$.
This divergence gives a term $\propto \ln(N)$, which we make explicit by writing
\begin{align}
    \sum_{n=1}^Ne_{n,\rm opt}\simeq \sum_{n=1}^\infty \left[e_{n,\rm opt}+\frac{1}{2\hat{\la}n}\right]-\frac{\ln N}{2\hat{\la}}
\end{align}
The constraint that the ribbon length in the optimal shape is $L$ reads $L/\ell-N_{\rm opt}=\ell^{-1}\sum_{\rm HC}[\int {\rm ds}-\ell]$, and up to the second order in $f_n$ :
\begin{align}
&L/\ell-N=    \demi\sum_{n=1}^{N_{\rm opt}}\left(\frac{\De f_n}{\ell}\right)^2\int_0^1{\rm d}u g_{n,\rm opt}'(u)^2\sim \mcal{S}\label{ribbonlengthequation}\\
\mcal{S} &=       \frac{1}{2\eta^2}\sum_{{n}=1}^{\infty} \frac{1}{\kappa_n^5}\left[-\tfrac{\kappa_n}{2}\tanh^2(\tfrac{\kappa_n}2)+3\tfrac{\kappa_n}2-3\tanh(\tfrac{\kappa_n}2)\right]\label{auto}
\end{align}
where the summation has been pushed to $+\infty$ since the series is convergent. Note that $\mcal{S}=\pa_\La\sum (e_{n,{\rm opt}}+1/(2\hat{\la}n))$, which is understood by considering Eq. \myref{pouet} : The derivative of this expression with respect to $\La$ has two parts, one implicit because $f$ depends on $\La$, which is identically zero because a first order variation of the optimal $f$ is always zero whatever its value, and one explicit because $\kappa_n^2$ depends on $\La$, which gives exactly the second-order approximation to $L/\ell-N$, i.e. $\mcal{S}$. Note also that we discarded the term $\propto \ln(N)$ which induces negligible corrections for long enough ribbons. Finally, one uses \myref{RN}, \myref{DeltaEad} and \myref{ribbonlengthequation} to write 
\begin{align}
    \Delta\mcal{E}/\ga\ell=\sum_{n=1}^\infty \left[e_{n,\rm opt}+\frac{1}{2\hat{\la}n}\right]+\frac{\hat{\la}}{2}\mcal{S}^2+(\sqrt{3}-\La)\mcal{S}
\label{nihao}\end{align}
This expression has to be optimized with respect to $\La$, a step equivalent to finding the optimal number $N_{\rm opt}$ of HC spanned by the ribbon at equilibrium (via Eq. \myref{ribbonlengthequation}). The fact that $\pa_\La \sum_{n}[e_{n,\rm opt}+1/(2\hat{\la}n)]=\mcal{S}$ shows that the equation fulfilled by the optimal $\La$ is either $\pa_\La\mcal{S}=0$ (which is impossible because $e_{n,\rm opt}$ is a convex function of $\La$) or
\begin{align}
    \frac{\La-\sqrt{3}}{\hat{\la}}=\mcal{S}\label{S}
\end{align}

To go further, one can evaluate numerically the preceding expression and solve for $\La$, but one can also make use of the fact that $\eta\gg 1$. For small values of $\hat{\la}/\eta=\la g\ell^3/\al$, the parameter $\kappa_n$ is slowly varying with $n$, and we can use the Euler-MacLaurin formula and replace the sum in \myref{auto} with an integral  (plus the correction terms ``$\frac{f(1)}{2}-\frac{f'(1)}{12}$'' \cite{DLMF} to be comprehensive at the order $\eta^{-4}$): After a cumbersome calculation, one finds
\begin{align}
    \La&=\sqrt{3}+\frac{1}{24\eta}-\frac{\sqrt{3}}{240\eta^2}+\frac{11-\frac{17}{48}\hat{\la}^2}{10080\eta^3}+o(\eta^{-3})\label{1EmcL}
\end{align}
It is remarkable that the first three orders of the expansion do not depend on $\hat{\la}$. This expression corresponds to the limit of large $\eta$ at constant $\hat\la$, and can be used also in the experimentally relevant limit of large $t$, with $\eta\propto t^3$ and $\hat{\la}\propto t$ : The term $\propto \hat{\la}^2\eta^{-3}$ must in this case be considered as $O(t^{-7})$ and overcomes the $O(\eta^{-3})=O(t^{-9})$ one. 

Another interesting case corresponds to $\hat{\la}=0$ whatever $\eta$. In this case, the lowest approximation for $\kappa_n$ in the case $\hat{\la}=0$ is $\kappa_n\sim 3^{1/4}/\sqrt{\eta}$, a result we obtained  in \cite{Jouanlanne2022} (formula (11) in \cite{Jouanlanne2022}, $\kappa_n\equiv 2\kappa$. The shape $g_n$ and the transverse amplitude $\De f_n$ are also the same).

Particularly interesting is the evolution of $\De f_n/\ell$ with ${n}=O(1)$ (lower part of the ribbon). In the approximation $\eta\gg 1$ we are considering here, $\kappa_n$ is for the lower part of the ribbon a small quantity and one can approximate \myref{deltaf} by
\begin{align}
    \frac{\sqrt{3}\De f_{n,\rm opt}}{\ell}&=\frac{1}{4\sqrt{3}\eta}\left[1-\frac{\sqrt{3}+\hat{\la}({n}-\demi)}{10\eta}\right]
\end{align}
As a result, a linear reduction of the lateral amplitude can be expected, with a slope $\hat{\la}/[40\sqrt{3}\eta^{2}]$ quite small with reasonable values of $\hat\ga$ and $\eta\gg 1$. This limit, $\eta\gg 1$, therefore does not appear as a relevant one to capture a signature of the gravitational effects. However, the theory is particularly clear. In the next paragraph, the same line of reasoning is followed in the opposite limit, when $\eta\ll 1$, a more complicated case due essentially to the fact that it is a expansion around a singular solution, reminiscent for instance of the zero temperature expansion of fermionic gas.

\subsection{Floppy ribbon limit : $\eta\ll 1$}\label{etapetit}

\subsubsection{The $\eta=0$ case}\label{B1}

In this opposite limit, the gravitational pull and the elastocapillary interaction oppose to each other since the capillarity  dominates largely the forces opposing the bending and forces the ribbon to a winding shape. To tackle this case, it is useful to note that $\hat\la\propto t$ ($t$ is the ribbon thickness), whereas $\eta\propto t^3$, which suggests that one considers first the case where formally $\eta=0$ and $\hat\la\neq 0$. We  define
\begin{align}
    \om_n^2&=\hat{\la}({n}-\demi)+{\La}\label{omn}
\end{align}
instead of $\kappa_n^2=\om_n^2/\eta$. The energy terms now read
\begin{align}
    e_n^{\circ}&=\om_n^2\int_0^\ell\frac{{\rm d}z}{\ell}[\sqrt{1+[f'(z)]^2}-1]-\frac{\De f_n}{\ell}
\end{align}
The optimal shape is readily $g^{\circ}(u)=u$, from which we deduce the optimal amplitude $\De f_n^\circ$ :
\begin{align}
\om_n^2\frac{\De f_n^\circ/\ell}{\sqrt{1+(\De f_n^\circ/\ell)^2}}-1=0 \Rightarrow \frac{\De f_{n}^{\circ}}{\ell}=\frac{1}{\sqrt{\om_n^4-1}}\label{deltaf0}
\end{align}
and the optimal energy $e_n^{\circ}=\sqrt{\om_n^4-1}-\om_n^2$. Once again the series in \myref{DeltaEad} has a logarithmic divergence, and we write in the large $L/\ell$ limit:
\begin{align}
    \De \mcal{E}^{\circ}/[\ga\ell]&\simeq\sum_{{n}=1}^\infty \left[e_n^{\circ}+\frac{1}{2\hat{\la}{n}}\right]-\frac{1}{2\hat{\la}}\ln(N)+\mcal{R}_N
    \end{align}
One finds the same type of calculations as before, and the ribbon length equation \myref{ribbonlengthequation} writes here
\begin{align}
    \frac{{\La}-\sqrt{3}}{\hat{\la}}&=\sum_{{n}=1}^\infty \frac{\om_n^2-\sqrt{\om_n^4-1}}{\sqrt{\om_n^4-1}}\label{lambda0}
\end{align}
where the series in the rhs is convergent. This equation is the self-consistent equation setting the value of $\La$ (present in the $\om_n$ as well). It is important to note (regarding the next subsection) that as for the $\eta\gg 1$ case, $\pa_\La (\sum_n e_n^\circ)=L/\ell-N$, which yields a similar structure between Eqs. \myref{S} and \myref{lambda0}.

The limit $\hat\la\rightarrow 0$ is instructive. The right hand side series is found (using the Euler-MacLaurin formula) equivalent to $[{\La}-\sqrt{{\La}^2-1}]/\hat{\la}$,   which gives ${\La}\sim 2$ in this limit. This is coherent in  
\myref{deltaf0} with the value $\Delta f_n^{\circ}/\ell=1/\sqrt{3}$ expected everywhere in this limit $\hat{\la}=\eta=0$.

\subsubsection{$\eta\ll1$ is a singular expansion }

To account for the cases where $\eta\ll 1$, but nonzero, the idea is to expand from the preceding solution. A key point is that the preceding solution is piecewise linear and therefore not twice differentiable as expected for a ribbon with a finite bending modulus. As a result, a naive expansion is bound to fail since the perturbative term should ``cure'' the zeroth-order singularity to yield a regular solution, which is just impossible.

To tackle this difficulty, one writes formally $f_n(z)=\frac{\De f^\circ_n}{\ell}z+\xi_n(z)$, where $\xi_n$ is a small departure from a zeroth-order which is similar to that computed in section \ref{B1},  but for the value $\La$, {\em which is not fixed at the zero-th order by \myref{lambda0}} but will be later on. One makes a Taylor expansion of the energy up to the second order in \myref{ousma} and \myref{ousmakappa}, what gives
\begin{multline}
    e_n=e_n^\circ\\+\frac{\eta\ell}{2}\frac{\om_n^2(\om_n^4-1)^{1/4}}{(\om_n^4-1)^{3/4}+1}\left[\int_0^\ell{\rm d}z[(\xi_n'')^2+k_n^2(\ell^{-1}\xi_n')^2]\right]
\label{formulevelue2}
\end{multline}
where \begin{align}
    k_n^2&=\frac{1}{\eta}\frac{(\om_n^4-1)^{5/4}[(\om_n^4-1)^{3/4}+1]}{\om_n^6}
\end{align}
and wherein $\om_n^2$ the value of $\La$ is yet to be determined.

\smallskip

At this point, it is crucial to express \myref{formulevelue2} back in terms of $f_n$ and optimize the energy with respect to $f_n$ rather than $\xi_n$ (a path also followed in \cite{Jouanlanne2022}). The reason is that the expected solution is at least $C^1$ and even twice differentiable, and that the zero-th order of our expansion is singular with respect to this requirement. As a result, as we cannot expect to recover with $\tfrac{\De f_n^\circ}{\ell}z+\xi_n(z)$ a twice differentiable solution, since $\xi_n(z)$ would be twice differentiable as the result of an optimization process, it is compulsory to reexpress $\xi_n$ in \myref{formulevelue2} in terms of $f_n$ and do the optimization directly on $f_n$. It is interesting to note that no obvious alternate method seems at hand to solve this issue and produce a first order expansion  in $\eta$ of the solution with the required regularity. This singularity will be manifest in the fact that the first correction to $\De f^\circ_n$ in $\De f$ will be $O(\sqrt{\eta})$, not $O(\eta)$.

Owing to these remarks, one write \myref{formulevelue2} making the substitution $\xi'_n=f_n'-\De f^{\circ}_n/\ell$ and $\xi_n''=f_n''$, to obtain
\begin{multline}
    e_n=e_n^\circ\\+\frac{\eta\ell}{2}\frac{\om_n^2(\om_n^4-1)^{1/4}}{(\om_n^4-1)^{3/4}+1}\left[\int_0^\ell{\rm d}z[(f_n'')^2+k_n^2(\ell^{-1}f_n')^2]\right]\\
    +\frac{1}{2\om_n^4}\left[\sqrt{\om_n^4-1}-2(\om_n^4-1)\frac{\De f_n}{\ell}\right]\label{formulevelue3}
\end{multline}
Comparison between \myref{formulevelue3} and \myref{pouet} shows that $f_n(z)=\De f_n h_n(u=z/\ell)$ where $h_n$ is given by the same formula as for $g_n$ in \myref{gn} but with $k_n$ replacing $\kappa_n$. Likewise, the amplitude $\De f_n$ minimizes 
  \begin{multline}
    e_{n,\rm opt}-e_n^\circ=\frac{1}{2}\frac{k_n(\om_n^4-1)^{3/2}}{\om_n^4(k_n-2\tanh\frac{k_n}{2})}\left(\frac{\De f_n}{\ell}\right)^2\\
    +\frac{1}{2\om_n^4}\left[\sqrt{\om_n^4-1}-2(\om_n^4-1)\frac{\De f_n}{\ell}\right]
\label{formulevelue}
\end{multline}
whence one gets
\begin{align}
\frac{\De f_{n,\rm opt}}{\ell}   &=\left[1-\frac{\tanh\frac{k_n}{2}}{k_n/2}\right]\left(\frac{1}{\sqrt{\om_n^4-1}}\right) \label{correction}
\end{align}
and
\begin{align}
    e_{n,{\rm opt}}=e_n^\circ
    +\frac{\sqrt{\om_n^4-1}}{\om_n^4k_n}\tanh\frac{k_n}{2}
\end{align}
which shows that the correction to the energy is consequently a convergent series. In \myref{correction}, $\La$ (implicit in $\om_n^2$ and $k_n$) is given by taking into account the constraint equation expressing that the length of the ribbon is $L$ : Up to the first order in $\xi_n'=f_n'-\De f_n^\circ/\ell$, this is
\begin{multline}
    \mcal{S}^\dag(\La)=\frac{L}{\ell}-N\\
    =\sum_n\left[\frac{\om_n^2-\sqrt{\om_n^4-1}}{\sqrt{\om_n^4-1}}-\frac{1+3\om_n^4}{4\om_n^6\sqrt{\om_n^4-1}}\frac{\tanh\frac{k_n}{2}}{k_n/2}\right]\\
    +\sum_n   \frac{\sqrt{\om_n^4-1}}{4\om_n^6}\left[1-\tanh^2\frac{k_n}{2}\right] \label{contraintelongueur}
\end{multline}
The value of $\La$ is finally found as the one minimizing
\begin{align}
    \frac{\De\mcal{E}(\La)}{\ga\ell}&=\sum_{n=1}^\infty \left[e_{n,\rm opt}+\frac{1}{2\hat{\la}n}\right]+\frac{\hat\la}{2}{\mcal{S}^\dag}^2+(\sqrt{3}-\La)\mcal{S}^{\dag}\label{pouet2}
\end{align}
which is simply done using for instance Matlab. It is worth noting that here the structure of the equation is more complicated than before ($\eta\gg 1$ or $\eta=0$ cases). This is due to the fact that in \myref{formulevelue3}, the explicit (i.e. not in $f_n$) dependence of the expression with respect to $\La$ is not of the form $\La(L/\ell-N)$ (as before), but involves also the curvature energy term. Consequently, we have no longer $\pa_\La\sum_n (e_{n,\rm opt}+1/(2\hat{\la}n))$  equating $\mcal{S}^{\dag}$, and \myref{pouet2} does not boil down to   $(\La-\sqrt{3})/\hat{\la}=\mcal{S}^{\dag}$ anymore.

\section{Comparison with experiments}

\subsection{Materials and Methods}

 A bubble column is produced by bubbling air via a pressure controller through a nozzle in a detergent solution (composed of water with 4.5 vol\% Fairy liquid, 1.5 vol\% Glycerol (Sigma-Aldrich) and 10 g.L$^{-1}$ J-Lube lubricant (Jorgensen Labs)). Once the square-cross section tube of width 15 mm is filled with bubbles, a polydimethylsiloxane thin ribbon (PDMS Sylgard 184, Dow Corning, 10:1 base to curing agent ratio, cured at 60°C) of width 14.5 $\pm$ 0.1 mm, hydrophilised via plasma cleaning treatment on both sides, is inserted in the central Plateau border. All details of the preparation procedure can be found in our previous article \cite{Jouanlanne2022}, and are visually highlighted in the video provided in the Supplementary Information. The surface tension of the solution is $\gamma=26 \pm 1$ mN/m, the thickness of the ribbons are varied using different spin-coating speeds ($t$=45, 55 and 127 $\unit{\micro\meter}$), and the elastic material parameters are taken as $E=1.7 \pm 0.2$ MPa for the Young's modulus, 0.45 for the Poisson's ratio, and $\rho=1027$ kg.m$^{-3}$ for the mass density.

 A specific care has been given in the current work to avoid any 3D effect in the case of long ribbons. To do so, the top of the ribbon has been pierced with six holes of 2 mm diameter, equally spaced on a horizontal line, to release lateral contraints at the upper clamping point. We also noticed that when the bottom of the ribbon arrives right at a node of the bubble column, there can be a pinning force exerted on the system, so we avoided this case and considered systems where the end of the ribbon is not at the extremity of a half-cell.

 The structure is allowed to drain for 30 minutes, to avoid motion artefacts, and is then scanned with a X-ray microtomograph EasyTom 150/160 from RX Solutions. A helical scan of 48 minutes is performed at resolution 12 $\unit{\micro\meter}$ to capture the deformation of the ribbon. The 3D image is then converted into 100 slices equally spaced, taken on planes perpendicular to the ribbon. On each slice, the amplitude of the deformation $\Delta f$ and the half-cell size $\ell$ is measured, and values are averaged over the 100 slices.

\subsection{Results and discussion}

\begin{figure}[H]
    \centering
    \includegraphics[width=0.5\textwidth]{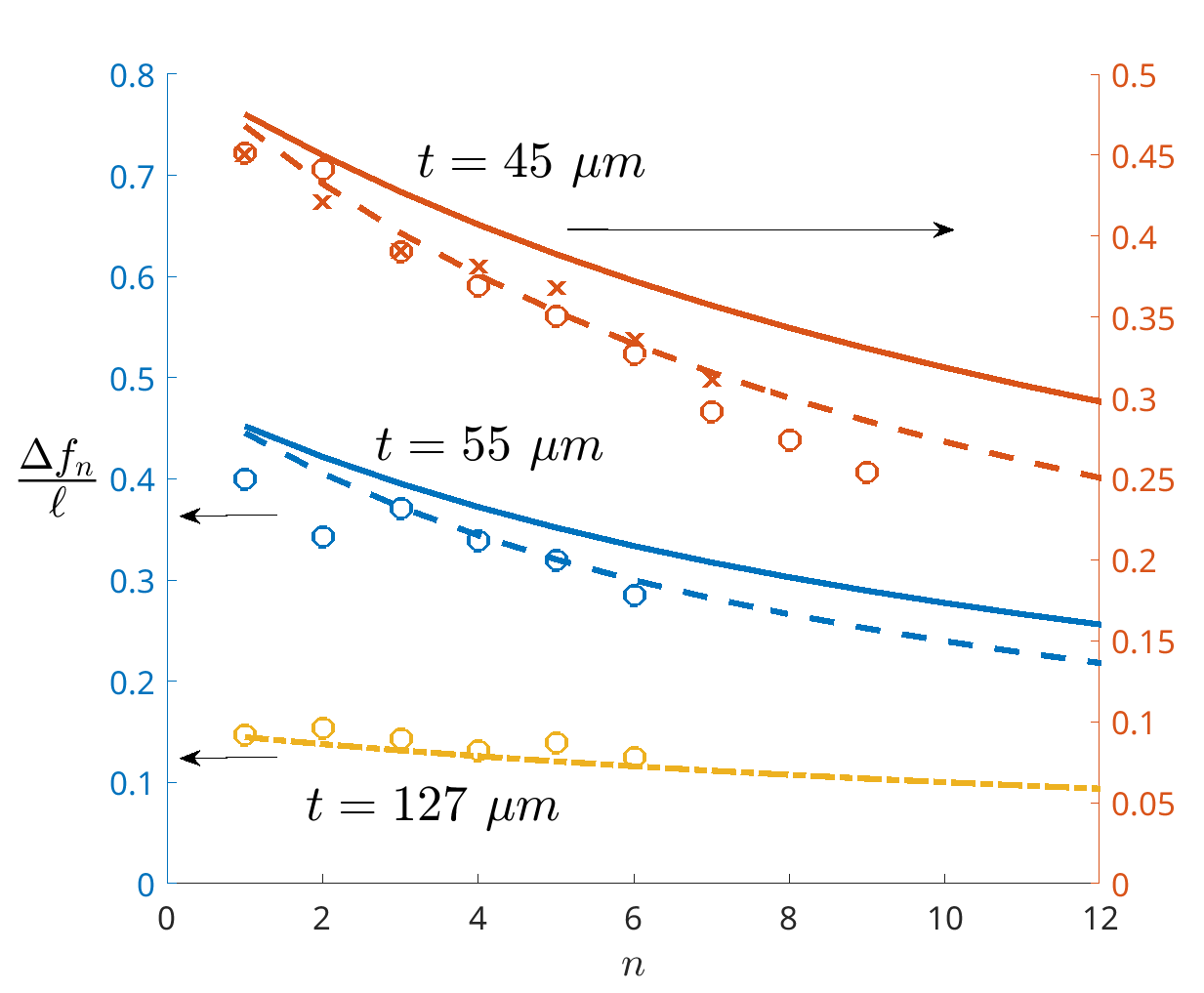}
    \caption{Decrease of $\De f_n/\ell$ as a function of $n$ for three different ribbon thicknesses : $t=45$ $\unit{\micro\meter}$ (red), $t=55$ $\unit{\micro\meter}$ (blue), and $t=127$ $\unit{\micro\meter}$ (yellow). The symbols are results of measurements from experiments (two different samples for $t=45$ $\unit{\micro\meter}$), the solid line are the result of the theory of section \ref{etapetit} ($\eta\ll 1$) without adjustable parameters, the dashed lines the theory where the dimensionless mass density $\hat{\la}$ has been adjusted (with a factor 1.4 and 1.5 for the red and blue curves respectively). The yellow dash-dotted curve is result of the theory of section \ref{etagrand} ($\eta\gg 1$). Notice that for sake of readability, the red curves $y$-axis is located to the right, whereas the other's is to the left. For the three experiments, the average values of $\ell$ are respectively (in mm) 5.7, 6.3 and 6.0 and the associated $\eta$ are $\num{1.9e-2}, \num{2.8e-2}$ and $0.38$.}
    \label{fig:results}
\end{figure}
We compared the prediction of the above theory with experiments, and the results are plotted in Figure \ref{fig:results}, where three different ribbons with different thicknesses were tested. For the thicknesses $t=45$  $\unit{\micro\meter}$ and $t=55$ $\unit{\micro\meter}$, the parameters $\eta$ are respectively  $\num{1.9e-2}$ and \num{2.8e-2} and it is relevant to use the theory of paragraph \ref{etapetit}. Interestingly, the theory predicts correctly $\De f_1/\ell$, but underestimates slightly the decrease of $\De f_n/\ell$, which is correctly taken into account only if the dimensionless mass density $\hat{\la}$ is allowed to be renormalized by a factor $1.4-1.5$. This discrepancy is probably due to the fact that the ribbon could be coated by 2 thin liquid layers of thickness of order $t_w\sim 10$ $\unit{\micro\meter}$ each. The viscoelastic nature of the detergent liquid used allows long-lasting structures but also slows down drainage, which makes this hypothesis likely. In the case of a thicker ribbon $t=\qty{127}{\micro\meter}$,  we used the theory of section \ref{etagrand} ($\eta\gg 1$), which captures well the experimental data, as shown in yellow in Figure \ref{fig:results}.

\section{Conclusion}

In this work, we have tackled the question of the equilibrium shape of elastic ribbons in 2D bubble columns, in the case where gravity, capillarity and elasticity compete to set the equilibrium architecture.  We have performed the analysis in multiple limit cases, and highlighted the most relevant one (floppy ribbon limit, for which gravity, capillarity and elasticity all significantly matter in the shape optimisation). This specific problem is fundamentally interesting as it involves an optimisation around a singular shape, provided by Plateau's laws when bubbles only are present.  Finally, we have successfully compared our analysis to experiments with multiple ribbon thicknesses, providing evidence of the importance of considering gravity in such systems and assessing the validity of our modeling.

\section{Acknowledgements}

We are grateful to Wiebke Drenckhan, Thierry Charitat and François Schosseler for fruitful discussions, to Damien Favier for his help in some preliminary tomography experiments, to ICS PECMAT team and especially Fouzia Boulmedais for the access to a spin-coater tool and to a plasma cleaner, and to Christophe Lambour for the fabrication of the square-section tube. We also thank Capucine Loth and Thibaut Schutz for their help in the recording of the Supplementary Material video. This work of the Interdisciplinary Institute HiFunMat, as part of the ITI 2021-2028 program of the University of Strasbourg, CNRS and Inserm, was supported by IdEx Unistra (ANR-10-IDEX-0002) and SFRI (STRAT’US project, ANR-20-SFRI-0012) under the framework of the French Investments for the Future Program. We also acknowledge funding from the IdEx Unistra framework (A. Hourlier-Fargette), and from the ANR (FOAMINT project, ANR-23-CE06-0014-01).

This research was funded, in whole or in part, by the ANR. A CC-BY public copyright license has been applied by the authors to the present document and will be applied to all subsequent versions up to the Author Accepted Manuscript arising from this submission, in accordance with the grant’s open access conditions.\\ 

\section{Appendix: Justification of the approximation $\ell$=constant}

 The volume variations induced by the ribbon are tiny and negligible, and $\ell={\rm const.}$ is a good approximation, provided the experimental setup warrants the generation of identical bubbles. To see this, one balances the deformation work done on a bubble by the typical energy of curvature of a zigzagging ribbon : $\de V/V\simeq \al/P_0V$, where $V$ is  a bubble volume (the geometry of the staircase bubbles is such that all relevant typical length scales, either direct or transverse, have the same order of magnitude $\ell$). With typical values for $(\al,P_0,V)$, one gets $\de\ell/\ell\sim \de V/V\sim \num{e-5}$, which validates the constant $\ell$ assumption.
Another even more important source of volume variation is associated to the fact that the ribbon thickness is not entirely negligible with respect to the tube lateral size : for a ribbon of size $\sim \qty{100}{\micro\meter}$, the induced (negative) volume variations are a priori $\de V/V\sim \num{e-2}$. If the pressure stays constant, $\de V$ must actually vanish, which would occur via a (positive) variation $\de\ell/\ell\sim \num{e-2}$  between a HC with a ribbon and a HC without. If the pressure actually increases slightly, $\de \ell/\ell$ is even smaller. All in all, we remark that $\de\ell/\ell$ is always quite small and we will henceforth neglect its variations.

\bibliography{apssamp}

\end{document}